\documentclass[useAMS,usenatbib]{mn2e}

\usepackage{epsfig}
\usepackage{amsmath}
\usepackage{natbib}
\usepackage{times}
\usepackage{amsfonts}
\usepackage{aas_macros}

\title[Non-Gaussianity and Forecasts]{The Impact of Non-Gaussianity upon Cosmological Forecasts}
\author[A. Repp, I. Szapudi, J. Carron, and M. Wolk]{A. Repp$^{1}$, I. Szapudi$^{1}$, J. Carron$^{1,2}$, and M. Wolk$^{1}$\\
$^{1}$Institute for Astronomy, 2680 Woodlawn Dr., Honolulu, Hawaii 96822, USA\\
$^{2}$Department of Physics and Astronomy, University of Sussex, Brighton BN1 9QH, U.K.}

\begin{document}

\date{\today}

\pagerange{\pageref{firstpage}--\pageref{lastpage}} \pubyear{0000}

\maketitle

\label{firstpage}

\begin{abstract}
The primary science driver for 3D galaxy surveys is their potential to constrain cosmological parameters. Forecasts of 
these surveys' effectiveness typically assume Gaussian statistics for the underlying matter density, despite the fact that the actual
distribution is decidedly non-Gaussian. To quantify the effect of this assumption, we employ an analytic expression for the
power spectrum covariance matrix to calculate the Fisher information for BAO-type model surveys.
We find that for typical number densities, at $k_\mathrm{max} = 0.5 h$ Mpc$^{-1}$, Gaussian assumptions
significantly overestimate the information on all parameters considered, in some cases by up to an order of magnitude. However, after marginalizing over a six-parameter set, the form of the covariance matrix 
(dictated by $N$-body simulations) causes the majority
of the effect to shift to the ``amplitude-like'' parameters, leaving the others virtually unaffected. We find that Gaussian assumptions at such wavenumbers can underestimate the dark energy
parameter errors by well over 50 per cent, producing dark energy figures of merit almost 3 times too large. Thus, for 3D galaxy surveys probing the non-linear regime, proper consideration of non-Gaussian effects
is essential.
\end{abstract}

\begin{keywords}
surveys -- cosmological parameters -- cosmology: theory
\end{keywords}

\section{Introduction}
\label{sec-intro}

Cosmological surveys provide the primary data set for characterizing the universe as a whole. It was
surveys of the cosmic microwave background (CMB) which revealed the small anisotropies \citep{COBE, WMAP, PlanckOverview} presumed
to be the progenitors of today's large scale structure; 
and a primary goal of recent galaxy surveys is to constrain the parameters of the
$\Lambda$-Cold Dark Matter ($\Lambda$CDM) model (e.g., \citealp{Tegmark2004} for the Sloan
Digital Sky Survey and \citealp{Cole2005} for the Two-degree Field Galaxy Redshift Survey). The term ``precision cosmology'' reflects
the resulting increased knowledge of these parameter values.

Inflationary theories typically predict the high degree of Gaussianity displayed by the CMB (e.g., \citealp{Bardeen1986, BondEfstathiou1987}). Given that the universe's matter distribution arose from the fluctuations visible in the CMB,
the galaxy power spectrum constitutes a powerful summary statistic for much of the information
in galaxy surveys (e.g., \citealp{Peebles1980, BaumgartFry1991, Martinez2009}).

Future widefield surveys will attempt to ascertain more precisely the galaxy power spectrum and thus the dark energy equation of state
\citep{DES, Euclid}. The planning of such surveys depends heavily on forecasting tools, which calculate the degree
to which the survey will constrain the parameters of interest. The primary such tool is the Fisher information matrix \citep{Fisher1925}, first
applied to cosmology by \citeauthor{Jungman1996a}
\citeyearpar{Jungman1996a, Jungman1996b}, by \citeauthor{Vogeley1996}
\citeyearpar{Vogeley1996}, and by \citeauthor{Tegmark+1997} \citeyearpar{Tegmark+1997}.
Calculating the Fisher matrix requires one to assume a particular form for the galaxy power spectrum covariance
(see Section~\ref{sec-method}); thus it also necessitates an understanding of the distribution of the underlying field.
One typically assumes this field to be Gaussian (e.g., \citealp{SeoEisenstein2007, Wang2010, JDEM}). For CMB fluctuations this assumption
is justified. However, for galaxy surveys, the over- and underdensities are distinctly non-Gaussian, for two reasons. First, galaxies' discrete
realization of the underlying dark matter field introduces Poisson shot noise into surveys (e.g., \citealp{Peebles1980, NSS2011}). 
Second -- and more problematically -- gravitational amplification of the primordial fluctuations alters the original
Gaussian matter field to a distinctly non-Gaussian distribution \citep{FryPeebles1978, Sharp1984, Bouchet1993, Szapudi1992,
Gaztanaga1994a}. 

As a result of this non-linearity, the different Fourier $k$-modes are now correlated rather than independent \citep{MeiksinWhite1999,
RimesHamilton2005, Neyrinck2006}. Hence, the power spectrum covariance matrix is no longer strictly diagonal (as in the Gaussian
case) but contains off-diagonal entries described by the trispectrum rather than by the power spectrum itself. Furthermore, surveys of
finite size have limited ability to resolve nearby $k$-modes, the resolving power being inversely proportional to the linear scale of the
survey. One effect of non-linear growth is to couple these unresolved nearby Fourier modes to the large-scale beat mode between
them; this is the ``beat coupling'' phenomenon, first described by \citet{HamiltonRS2006} (see also \citealp{Sefusatti2006}). Thus
the covariance of the non-linear power spectrum is dominated by the largest scale (super-survey) modes, and an adequate description
of the information in the non-linear regime must account for both the intra-survey trispectrum
and the coupled super-survey modes. Hence, a significant portion of the information inherent in the survey escapes
from the power spectrum, causing an ``information plateau'' at higher values of $k$ \citep{LeePen2008, Carron2011, CarronNeyrinck2012}.

Thus the (typical) forecasting assumption of a Gaussian field leads (at high $k$-values) to inaccurate information estimates and
therefore to inaccurate
forecasts. Existing work has not thoroughly investigated the quantitative impact of this assumption. We thus present here a method
for quantifying the effect which Poisson sampling of a non-Gaussian field exercises upon forecasts. Initial results show that at
$k_\mathrm{max}=0.5h$ Mpc$^{-1}$ it significantly influences information estimates for all parameters considered,
by up to an order of magnitude for certain amplitude-like parameters.

Previous work by \citet{TakadaJain2009} showed for weak lensing surveys that marginalization over a parameter set can reduce the effect of non-Gaussianity on forecasting. We show (for galaxy surveys) the reason that marginalization affects certain parameters more than others. In particular, it is the
amplitude-like parameters which carry the effect of non-Gaussianity after marginalization, leading in turn to a dark energy figure of merit up to 2.9 times too large.
Upon publication we will make available a code which realizes our method.

The structure of this paper is as follows. In Section~\ref{sec-method} we describe our
method of calculating Fisher information. In Section~\ref{sec-testing} we compare our method to simulations. In
Sections~\ref{sec-results-Fisher}--\ref{sec-results-errors} we quantify the impact of assuming Gaussianity, and we show that this
assumption results in underestimates of the dark energy parameter errors of more than 50 per cent. We discuss the effects
of marginalization in Section~\ref{sec-disc}, and we conclude in Section~\ref{sec-concl}.

\section{Method}
\label{sec-method}

The Fisher matrix quantifies the information which a random variable carries about the parameters upon which it depends. The 
random variable in this case is the power spectrum $P(k)$, which depends on a vector of cosmological parameters
$\boldsymbol{\theta}= [\theta_1, \theta_2,
\ldots, \theta_n]$. If $p(P(k); \boldsymbol{\theta})$ is the probability of observing values $P(k)$ of the power spectrum given a set
of parameter values $\boldsymbol{\theta}$, then the entries of the ($n \times n$) Fisher matrix are 
\begin{equation}
F_{ij} = \left<\frac{\partial\ln p(P(k);\boldsymbol{\theta})}{\partial \theta_i} \frac{\partial\ln p(P(k);\boldsymbol{\theta})}{\partial \theta_j} \right>,
\end{equation}
or equivalently
\begin{equation}
F_{ij} = -\left<\frac{\partial^2\ln p(P(k);\boldsymbol{\theta})}{\partial \theta_i \partial \theta_j} \right>.
\end{equation}

The Cram\'er-Rao Inequality \citep{Rao, Cramer} elucidates the importance of the Fisher matrix, stating that
if $\hat{D_i}$ is a set of unbiased estimators for observable quantities $D_i$, then 
\begin{equation}
\mathrm{Cov}(\hat{D})_{ij} \ge \sum_{\alpha, \beta} \frac{\partial D_i}{\partial \theta_\alpha} \left(F^{-1}\right)_{\alpha\beta} \frac{\partial D_j}{\partial \theta_\beta},
\end{equation}
where $\mathrm{Cov}(\hat{D})$ is the covariance matrix for the estimators $\hat{D_i}$. This inequality thus provides the link
between the Fisher matrix and the (co-)variances of the observables. Two results follow. First, if the estimated quantities $D_i$ are identical to the parameters $\theta_i$, then the inverse Fisher matrix sets a lower bound on the possible covariance of the parameters:
\begin{equation}
\Sigma_{ij} \ge (F^{-1})_{ij},
\label{eqn:CR}
\end{equation}
where $\Sigma$ is the covariance matrix for the parameters $\theta_i$.

Second, if we take the observables $D_i$ to be the power spectrum values $P(k_i)$ and further require that all information be derived from
these observables in the context of a physical theory, then the inequality becomes an equality, and 
\begin{equation}
F_{\alpha\beta} = \sum_{k_i, k_j \le k_\mathrm{max}} \left( \frac{\partial P(k_i)}{\partial\alpha} \mathrm{Cov}_{ij}^{-1} \frac{\partial P(k_j)}{\partial \beta} \right),
\label{eqn:Fdef}
\end{equation}
where $\alpha$ and $\beta$ are cosmological parameters of interest and where $\mathrm{Cov}_{ij}$ is the covariance matrix of the power
spectrum modes. It is through $\mathrm{Cov}_{ij}$ that the non-Gaussianity of the matter field enters the calculation.

Assuming no shot noise, Equation~\ref{eqn:Fdef} yields the following standard approximation \citep{Tegmark1997} for the Fisher matrix on a Gaussian field:
\begin{equation}
F^G_{\alpha\beta} = \frac{V}{2} \int^{k_\mathrm{max}}_{k_\mathrm{min}} \frac{dk\,k^2}{2\pi^2} \frac{\partial \ln P(k)}{\partial\alpha} \frac{\partial \ln P(k)}{\partial \beta}.
\end{equation}
Here $V$ is the volume of the survey, and $k_\mathrm{min}$ and $k_\mathrm{max}$ are the
minimum and maximum wavenumbers under consideration. In most cases it suffices
to assume that $k_\mathrm{min}$ vanishes.

Our goal is to contrast the information $F^G_{\alpha\beta}$ in a Gaussian field
with the information $F_{\alpha\beta}$ in a Poisson-sampled non-Gaussian field.
Since non-Gaussianity affects the Fisher
matrix through $\mathrm{Cov}_{ij}$, we require an expression for the power spectrum covariance
in the non-Gaussian case. \citealt{SSFPaper} (hereinafter CWS15) introduce the following approximation based on simulations by \citet{Neyrinck2011} and \citet{MohammedSeljak2014}:
\begin{equation}
\mathrm{Cov}_{ij} = \delta_{ij} \frac{2 (P(k_i) + \frac{1}{\overline{n}})^2}{N_{k_i}} + \sigma^2_\mathrm{min} P(k_i)P(k_j),
\label{eqn:CWS15}
\end{equation}
where $N_{k_i}$ is the number of Fourier modes in the $k_i$-shell in Fourier space, and
where $\overline{n}$ is the average galaxy number density.

The behavior of the Fisher information in the non-linear
regime is a direct consequence of this specific form for the covariance matrix. In this
expression, the first term is the Gaussian covariance, modified to include shot noise;
the second term parametrizes the non-Gaussianity of the underlying field
by means of $\sigma^2_\mathrm{min}$. As CWS15 note (see discussion around their
eqn.~29), one can obtain this covariance matrix by starting with a Gaussian field and then modulating
it with a stochastic amplitude parameter whose variance is $\sigma^2_\mathrm{min}$. As we note
later (see Section~\ref{sec-disc} and Appendix~\ref{sec:appendix}), the result is that the effects
of non-Gaussianity appear in the amplitude-like parameters, while the other parameters retain
essentially Gaussian behavior.

The quantity $\sigma^2_\mathrm{min}$ represents the minimum achievable variance for a log-amplitude parameter $\ln A_0$, defined by
\begin{equation}
\frac{\partial P(k_i)}{\partial \ln A_0} = P(k_i).
\label{eqn:lnA0}
\end{equation}
Note that this parameter $A_0$ is distinct from the initial amplitude $A$ of the linear power spectrum. $A_0$ measures non-linear amplitude, and one could thus take it to equal $\sigma_8^2$ in the linear regime, in which case it would differ slightly from $\sigma_8^2$ on the translinear scales which we consider.

$\sigma^2_\mathrm{min}$ thus marks an inherent information plateau for the standard power spectrum of a non-Gaussian field. CWS15 further
decompose $\sigma^2_\mathrm{min}$ into $\sigma^2_\mathrm{SS}$ and $\sigma^2_\mathrm{IS}$. The first component
$\sigma^2_\mathrm{SS}$ expresses the impact of large-scale (super-survey) modes through beat coupling. The second component
$\sigma^2_\mathrm{IS}$ expresses the impact of small-scale (intra-survey) couplings on the trispectrum.

CWS15 then derive the following expression for the Fisher matrix $F_{\alpha\beta}$:
\begin{equation}
F_{\alpha\beta} = F^G_{\alpha\beta} - \sigma^2_\mathrm{min} \frac{F^G_{\alpha\,\ln\!A_0} F^G_{\ln\!A_0\,\beta}}{1 + \sigma^2_\mathrm{min} F^G_{\ln\!A_0\,\ln\!A_0}}.
\label{eqn:SSF}
\end{equation}

To include Poisson sampling, we note that shot noise appears only in the Gaussian term of the covariance expression; thus it suffices to modify the expression for $F^G$ (following \citealp{Tegmark1997}) to read
\begin{equation}
F^G_{\alpha\beta} = \frac{V}{2} \int^{k_\mathrm{max}}_{k_\mathrm{min}} \frac{dk\,k^2}{2\pi^2} \frac{\partial \ln \left(P(k) + \frac{1}{\overline{n}}\right)}{\partial\alpha} \frac{\partial \ln \left(P(k) + \frac{1}{\overline{n}}\right)}{\partial \beta},
\end{equation}
with Equation~\ref{eqn:SSF} remaining unchanged.

The final step is to approximate $\sigma^2_\mathrm{min}$; we again follow CWS15. The hierarchical ansatz \citep{Peebles1980, Fry1984} allows us to approximate both
components analytically. We can write the intra-survey component as 
\begin{equation}
\label{eqn:ISint}
\sigma^2_\mathrm{IS} = 8\int_V \frac{d^3 x}{V} \int_V \frac{d^3 y}{V} \frac{\xi^2(x-y)}{\sigma^2},
\end{equation}
where $\xi$ is the two-point correlation function and $\sigma^2$ is the variance $\langle\delta^2\rangle$ of the matter field; we can further approximate Equation~\ref{eqn:ISint} as
\begin{equation}
\label{eqn:ISsimp}
\sigma^2_\mathrm{IS} = 8\frac{P(k_\mathrm{max})}{V}.
\end{equation}
Likewise, from \citet{TakadaHu2013} we can express the super-survey component as
\begin{equation}
\label{eqn:SSsimp}
\sigma^2_\mathrm{SS} = \left(\frac{26}{21}\right)^2 \cdot \frac{1}{2\pi^2} \int dk\, k^2 P_\mathrm{lin}(k) W(k)^2.
\end{equation}
In this expression, $P_\mathrm{lin}(k)$ specifies the linear power spectrum, and $W(k)$ is the Fourier transform of the window function
$W(x)$ (which equals $1/V$ inside the volume and vanishes outside of it). The factor of $26/21$ reflects the fact that we are defining
the fluctuations $\delta$ with respect to the observed local density (as appropriate for a galaxy survey). If we were defining it
with respect to a global density (e.g., for a weak lensing survey) the factor would become $68/21$. The factors differ because the local mean density includes the contribution of background modes, so that
\begin{equation}
\delta(x)_\mathrm{local} = \frac{\delta(x)_\mathrm{global}}{1 + \delta_\mathrm{bkgd}}.
\end{equation}
See
table~1 of \citet{WolkNeutrino} for values of $\sigma^2_\mathrm{IS}$ and $\sigma^2_\mathrm{SS}$
at redshifts from $z=0$ to 2. Note also that in Equations~\ref{eqn:ISsimp} and \ref{eqn:SSsimp}, $P(k)$ and $P_\mathrm{lin}(k)$ refer to \emph{unbiased} power spectra, so that $\sigma^2_\mathrm{min}$ does not depend upon the galaxy bias one assumes.

Since $\sigma^2_\mathrm{min}$ is simply the sum of $\sigma^2_\mathrm{IS}$ and $\sigma^2_\mathrm{SS}$,
we can now  calculate the Fisher information matrix under the assumption of Poisson sampling of
a non-Gaussian field.

To implement this formula, we generate (non-linear) power spectra with \texttt{CAMB}\footnote{http://camb.info/} (Cosmic Anisotropies in the Microwave Background, \citealp{CAMB}), which uses the halofit model of \citet{Takahashi2012}. From these spectra we numerically calculate partial derivatives using symmetric difference
quotients ($\Delta \alpha/\alpha = 0.05$). For fiducial cosmological parameter values we use Planck 2015 \citep{PlanckValues}.

\section{Validation}
\label{sec-testing}To validate this procedure, we attempted to reproduce the error-bar results
of \citealt{Neyrinck2011b} (hereinafter N11). N11 bases his calculations on the 37 cosmological simulations composing the Coyote Universe suite \citep{Coyote1, Coyote2, Coyote3}. He 
obtains an ensemble of power spectra for each cosmology, and from these power spectra
he estimates the covariance. Since we use Equation~\ref{eqn:CWS15} to approximate the covariance, comparison of our
results with these simulations will both validate our method in general and further establish the accuracy of this approximation.

Equation~\ref{eqn:Fdef} shows that calculation of Fisher information requires not only the covariance
but also the derivatives of the power spectrum. N11 estimates these derivatives directly from the variations
in the realizations of the Coyote Universe cosmologies. These cosmologies specify parameters
$\Omega_m h^2$, $\Omega_b h^2$, $n_s$, $w$, and $\sigma_8$, from which they calculate the Hubble constant parameter
$h$ (based on the angular scale of the CMB acoustic peaks, \citealp{Coyote3}).

We must however consider two subtleties. First, such a procedure makes $h$ a function of the other
five parameters rather than an independent variable. This fact mandates some care if we are to
reproduce N11's partial derivatives. If $\boldsymbol{\alpha}$
is the vector of the five parameter values, then
\begin{equation}
P(k) = P(\boldsymbol{\alpha}, h(\boldsymbol{\alpha}), k);
\end{equation}
thus a derivative calculated from the Coyote Universe simulations involves an extra
term due to the implicit $\boldsymbol{\alpha}$-dependence of $h$:
\begin{equation}
\left(\frac{\partial P}{\partial \boldsymbol{\alpha}}\right)_\mathrm{CU} = \frac{\partial P}{\partial \boldsymbol{\alpha}} + \frac{\partial P}{\partial h} \frac{\partial h}{\partial \boldsymbol{\alpha}}.
\end{equation}
We therefore must estimate $\partial h/\partial \alpha_i$ for given values of $\boldsymbol{\alpha}$. To do
so, we employ the parameter values from table~1 of \citet{Coyote3}; we form a simplex from the six points
closest to $\boldsymbol{\alpha}$ in 5-dimensional parameter space; and we determine $\partial h$ by
interpolation on that simplex.

The second subtlety concerns the fact that it suffices for N11 to model $\partial P(k)/\partial \boldsymbol{\alpha}$ in terms of fluctuations away from the mean power spectrum of the 37 cosmologies. This procedure yields one derivative for the Coyote Universe suite as a whole, whereas our method calculates a separate $\partial P(k)/\partial \boldsymbol{\alpha}$ for each cosmology. Thus we must compare N11's one error bar (for each parameter) to our ensemble of 37 error bars (for each parameter). 

\begin{figure}
    \leavevmode\epsfxsize=9cm\epsfbox{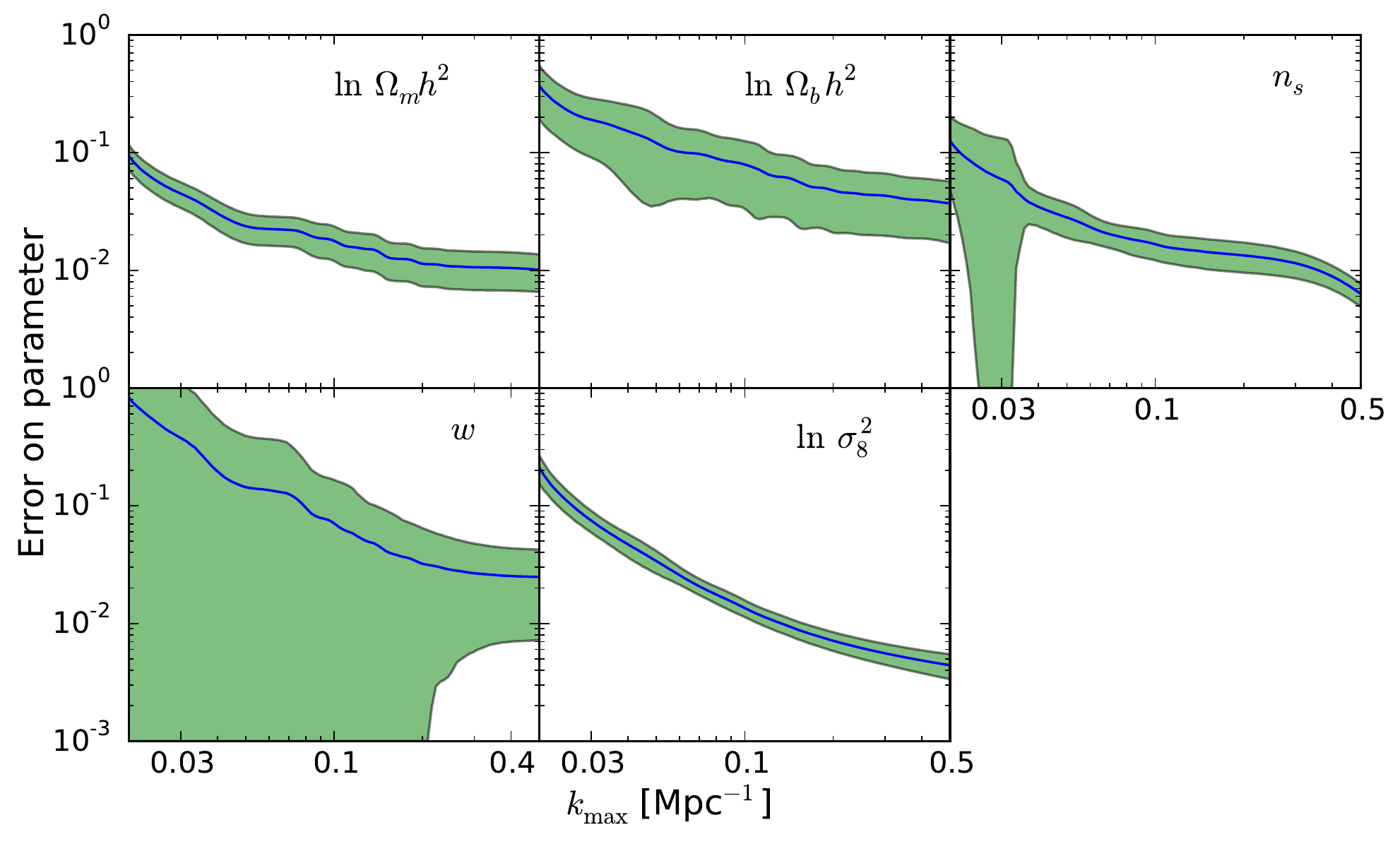}
    \caption{Size of (unmarginalized) error bars for the five Coyote Universe parameters. The blue line in each panel
    denotes the mean over the 37 Coyote Universe cosmologies, and the green shaded regions mark the range
    of 1-$\sigma$ variation over those cosmologies. Comparison with fig.~7 of N11 demonstrates that our results agree with the N11 simulation-based results within 1$\sigma$ for $k_\mathrm{max}$ = 0.1 Mpc$^{-1}$ and within 1.5$\sigma$ for $k_\mathrm{max}$ = 0.5 Mpc$^{-1}$, with one exception in both cases. The exception is $\ln \sigma_8^2$, which differs from N11's results by 1.5$\sigma$ in the first case and 2.4$\sigma$ in the second (see text for discussion). We conclude that our method returns results comparable to those obtained from simulations.}
\label{fig:Neyrinck}
\end{figure}

We hence calculate Fisher information (and thus error bars) as functions of $k_\mathrm{max}$
for each of the 37 Coyote Universe cosmologies, and we display the results in Fig.~\ref{fig:Neyrinck}. The blue line in each panel denotes the mean error bar over the 37-member ensemble, and the green shading indicates the 1-sigma range of the error bars (again over the suite of 37 cosmologies). The corresponding curves in N11 are those plotted in black along the diagonal of his fig.~7. Comparison at $k_\mathrm{max} = 0.1$ Mpc$^{-1}$ shows that, with one exception, our results agree with those of N11 to within 1$\sigma$ (where $\sigma$ here is the standard deviation of the error bar sizes over the Coyote Universe ensemble). At $k_\mathrm{max} = 0.5$ Mpc$^{-1}$ the results agree (again with one exception) to within 1.5$\sigma$. The exception in both cases is $\ln \sigma^2_8$, where N11's results differ from ours by 1.5$\sigma$ and 2.4$\sigma$ at $k=0.1$ and $0.5$ Mpc$^{-1}$, respectively. However, even in this more nearly discrepant case, the error bars we obtain for
$\ln \sigma^2_8$ in the non-Gaussian case are on average \emph{lower} than those which N11 obtain from simulations. Thus N11's
simulations show that the effect of non-Gaussianity on $\sigma_8$ will be at least as large as stated in this work -- and
possibly larger. Hence we conclude that our method, based on the CWS15 approximation, yields results comparable to cosmological
simulations.

For the remainder of this paper we fix $h=0.6774$ (the Planck 2015 value) for our calculations, and we quote results in terms of $h$ (e.g., units of $k$ are $h$ Mpc$^{-1}$).

\begin{figure*}
    \leavevmode\epsfxsize=18cm\epsfbox{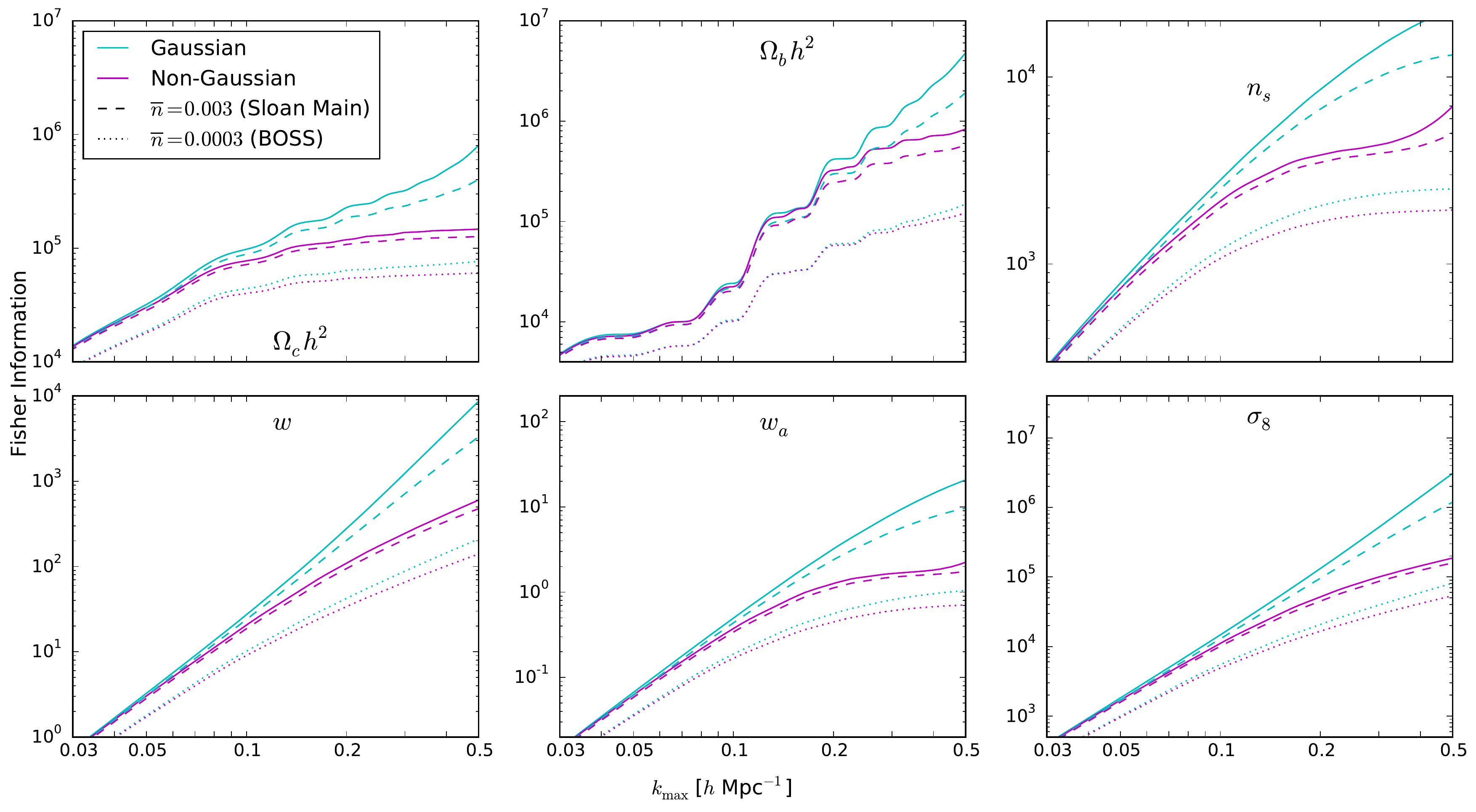}
    \caption{Fisher information under both Gaussian and non-Gaussian assumptions for a hypothetical 1-Gpc$^3$ survey at $z=0.5.$ Solid curves represent the limiting case of a continuous matter field; dashed and dotted curves assume Poisson sampling with $\overline{n} = 0.003$ and 0.0003$h^3$ Mpc$^{-3}$ respectively (comparable to the SDSS MGS and to the BOSS sample). We use Equation~\ref{eqn:SSF} to calculate the information for non-Gaussian cases. Note that the scale of the vertical axis differs from panel to panel. We assume a galaxy bias of unity.}
\label{fig:absinfo}
\end{figure*}

\begin{figure*}
    \leavevmode\epsfxsize=18cm\epsfbox{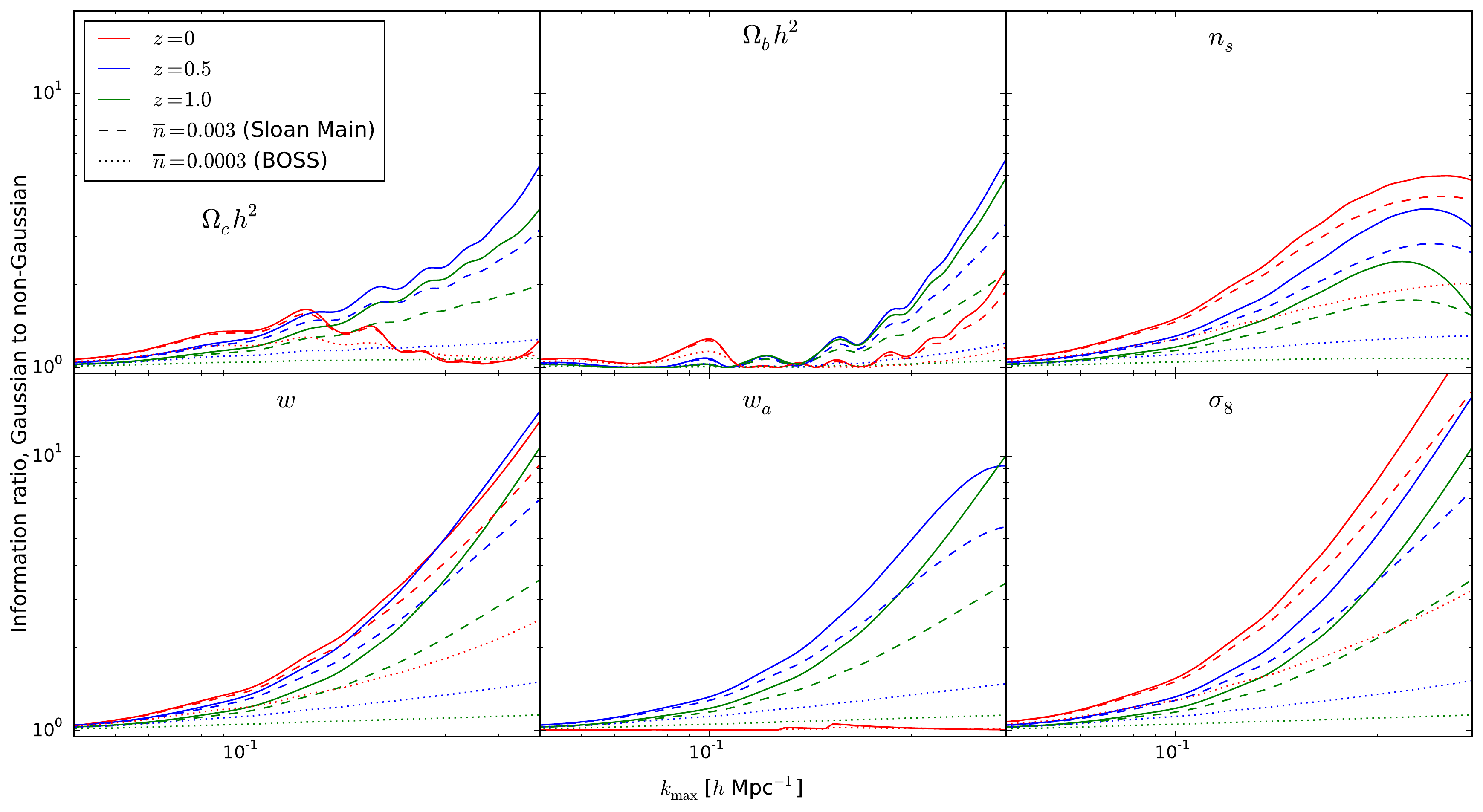}
    \caption{Ratio of Fisher information under Gaussian assumptions to that under non-Gaussian assumptions, for three hypothetical 1-Gpc$^3$ surveys at $z=0, 0.5$, and 1. Solid curves represent the limiting case of a continuous matter field; dashed and dotted curves assume Poisson sampling with $\overline{n} = 0.003$ and 0.0003$h^3$ Mpc$^{-3}$ respectively (comparable to the SDSS MGS and to the BOSS sample). We use Equation~\ref{eqn:SSF} to calculate the information for non-Gaussian cases. Note that for typical number densities, the assumption of Gaussian covariance can cause an order of magnitude overestimate for the information content on certain parameters. We assume a galaxy bias of unity.}
\label{fig:zplots}
\end{figure*}

\section{Results: Fisher Information}
\label{sec-results-Fisher}
We can now compare the Fisher information on various parameters, under different covariance assumptions (Gaussian vs. non-Gaussian), and under different sampling assumptions (continuous vs. Poisson). We here consider the following six parameters: $\Omega_c h^2$ and
$\Omega_b h^2$ (the physical cold dark matter and baryon densities, respectively); $n_s$ (the spectral index for the primordial power spectrum);
$w$ and $w_a$ (the dark energy equation of state and its derivative); and $\sigma_8$ (the linear amplitude parameter). We simulate three 1-Gpc$^3$ surveys, at $z=0,$ 0.5, and 1; and we assume a galaxy bias of unity. We consider two mean galaxy number densities throughout. The first is $\overline{n}=0.003h^3$ Mpc$^{-3}$, which corresponds roughly \citep{HuHaiman2003} to the Sloan Digital Sky Survey (SDSS) Main Galaxy Sample (MGS). The second is $\overline{n}=0.0003h^3$ Mpc$^{-3}$, which corresponds roughly \citep{Anderson2012} to the Baryon Oscillation Spectroscopic Survey (BOSS). For comparison we also consider the limiting case of a continuous field. 

We first calculate the absolute Fisher information for each parameter. As an example, Fig.~\ref{fig:absinfo} shows the information in the survey at $z=0.5$ (chosen since this redshift represents a regime of transition from matter domination to dark energy domination). One notes that the different covariance assumptions typically begin to affect the results even before $k_\mathrm{max}=0.1h$ Mpc$^{-1}$, although the details vary from parameter to parameter. 

Since our purpose is to quantify the impact of non-Gaussianity, it is more instructive to consider the ratio (Gaussian to non-Gaussian)
of information rather than the absolute amount of information. Thus we display these ratios in 
Fig.~\ref{fig:zplots} for each parameter, for each redshift, and for each sampling assumption.
In general the effects  increase with wavenumber, as expected given the information plateau described in Sections~\ref{sec-intro}
and \ref{sec-method}.

The magnitude of the effect is clearly non-negligible, especially for the dark energy parameters and for $\sigma_8$. Consider for instance a 1-Gpc$^3$ survey similar to the SDSS MGS, and assume that the survey considers wavenumbers up to $k=0.5h$ Mpc$^{-1}$. The lower left panel of Fig.~\ref{fig:zplots} shows that assuming Gaussian covariance at $z=0$ will lead one to predict an information content for $w$ more than 9 times greater than in the real (non-Gaussian) universe. Since Fisher information scales as volume, the 1-Gpc$^{3}$ survey will thus yield only the information expected from a 0.1-Gpc$^3$ survey. In other words, Gaussian assumptions can cause an order of magnitude overestimate of Fisher information.

For smaller number densities the effect can still be significant: reference to the lower right panel of Fig.~\ref{fig:zplots} shows that forecasts of information on $\sigma_8$ for a BOSS-type 1-Gpc$^3$ survey at $z=0$ are three times greater under Gaussian assumptions than under non-Gaussian at $k_\mathrm{max}=0.5h$ Mpc$^{-1}$. Thus, while it is safe to assume Gaussian statistics in the linear regime, doing so at higher wavenumbers can drastically inflate the predicted information content. Future surveys such as Euclid aim to probe this non-linear
region on scales approaching $1h$ Mpc$^{-1}$ \citep{Hearin2012}. Thus, though the effect is subtle for linear
wavenumbers, the effect will be profound for future surveys, as we demonstrate in the next two sections.

\section{Marginalization}
\label{sec-disc}

\begin{figure*}
    \leavevmode\epsfxsize=18cm\epsfbox{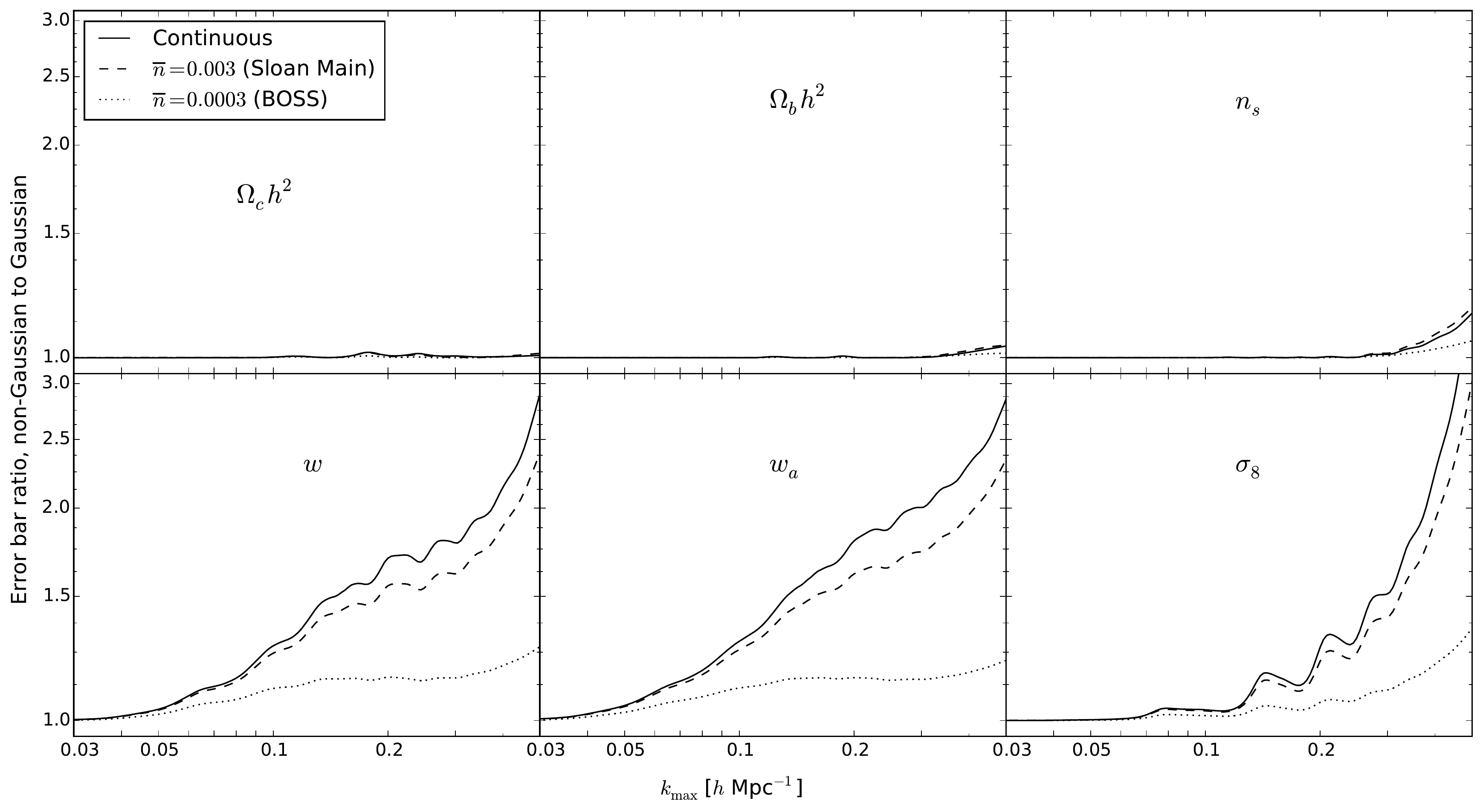}
    \caption{Ratio (non-Gaussian to Gaussian) of the sizes of marginalized error bars for the combined results of three hypothetical 1-Gpc$^3$ surveys at $z=0, 0.5$, and 1. Solid lines represent the limiting case of a continuous matter field; dashed and dotted lines assume Poisson sampling with $\overline{n} = 0.003$ and 0.0003 $h^3$ Mpc$^{-3}$, respectively (comparable to the SDSS MGS and to the BOSS sample). We use Equation~\ref{eqn:SSF} to calculate the information for non-Gaussian cases. We assume a galaxy bias of unity.}
\label{fig:margerr}
\end{figure*}

\begin{table}
  \leavevmode
  \caption{Error ratios, non-Gaussian to Gaussian}
  \label{tab:errors}
  \begin{tabular}{lcccccc} \hline 
  				& \multicolumn{5}{c}{Ratios of Error Bar Sizes} \\
$k_\mathrm{max}$ [$h$ Mpc$^{-1}$] & $\Omega_c h^2$	& $\Omega_b h^2$	& $n_s$	& $w$	& $w_a$	& $\sigma_8$\\ \hline
0.1							& 1.00			& 1.00			& 1.00	& 1.28	& 1.30	& 1.04\\
\hspace{12pt}$\overline{n} = 0.003$	& 1.00			& 1.00			& 1.00	& 1.25	& 1.27	& 1.03\\ 
\hspace{12pt}$\overline{n} = 0.0003$	& 1.00			& 1.00			& 1.00	& 1.11	& 1.11	& 1.02\\ \hline
0.2							& 1.01			& 1.00			& 1.00	& 1.71	& 1.81	& 1.30\\
\hspace{12pt}$\overline{n} = 0.003$	& 1.00			& 1.00			& 1.00	& 1.56	& 1.62	& 1.24\\
\hspace{12pt}$\overline{n} = 0.0003$	& 1.00			& 1.00			& 1.00	& 1.15	& 1.15	& 1.07\\ \hline
0.5							& 1.01			& 1.04			& 1.15	& 2.80	& 2.78	& 3.87\\
\hspace{12pt}$\overline{n} = 0.003$	& 1.01			& 1.04			& 1.17	& 2.34	& 2.30	& 2.88\\ 
\hspace{12pt}$\overline{n} = 0.0003$	& 1.01			& 1.01			& 1.05	& 1.26	& 1.21	& 1.33\\ \hline
\hline\
 \end{tabular}
\medskip
\\Ratio of error bar sizes under non-Gaussian assumptions to those under Gaussian assumptions. Calculations assume the combination of three hypothetical 1-Gpc$^3$ surveys at $z=0$, 0.5, and 1. For each maximum wavenumber the first row displays the limiting case of a continuous matter distribution, while the second and third rows assume Poisson sampling with $\overline{n} = 0.003$ and $0.0003h^3$ Mpc$^{-3}$ respectively (representative of the SDSS MGS and BOSS samples). We assume a galaxy bias of unity.
\end{table}

Fisher analysis in this context is a tool to forecast confidence limits for future surveys. Thus we must translate these information ratios into error bar sizes. In calculating unmarginalized errors for a single parameter, one assumes perfect knowledge of the remainder of the parameter set; thus one considers only the diagonal elements of the Fisher matrix. So in the unmarginalized case, the minimum standard deviation for the $i$th parameter is
\begin{equation}
\sigma_i = 1/\sqrt{F_{ii}}
\end{equation}
by Equation~\ref{eqn:CR}. Inspection of Fig.~\ref{fig:zplots} shows that the ratio of unmarginalized error bars can be greater than 3 for parameters like $\sigma_8$.

More useful for our purposes, however, are the error bars derived from marginalizing over the possible
values of the remaining parameters. In this case one must invert the entire Fisher matrix to obtain the
matrix of parameter covariances (Equation~\ref{eqn:CR}). We show in Appendix~\ref{sec:appendix} that if the amplitude parameter
$\ln A_0$ (defined by Equation~\ref{eqn:lnA0}) is part of the parameter
set, and if the Fisher matrix is invertible, then the entire effect of non-Gaussianity appears in the error bars for $\ln A_0$;
perhaps surprisingly, non-Gaussianity has no effect on the remaining parameters. This result is a direct consequence of the
form of the power spectrum covariance matrix derived from simulations and expressed in Equation~\ref{eqn:CWS15}.

However, there are two caveats to this result. First, it depends on the invertibility of the Gaussian Fisher
matrix and otherwise fails. Furthermore, as $F^G$ approaches singularity, the matrix becomes increasingly
ill-conditioned and thus numerically unstable with respect to inversion. In such a case one is
attempting to analyse too many parameters simultaneously. One solution is to shrink the parameter
set; another is to impose a prior; a third is to obtain more data to break the degeneracy (for instance, by combining data from multiple redshifts).

The second caveat is that no standard cosmological parameter is strictly equal to $\ln A_0$; thus one cannot in practice expect all non-Gaussian effects to accumulate in one parameter. However, one does expect ``amplitude-like'' parameters (such as $\sigma_8$ and the
dark energy parameters $w$ and $w_a$) to to be the ones which particularly exhibit the effects of non-Gaussianity.

Thus, if $\ln A_0$ is part of the parameter set, marginalization concentrates the entire effect of the non-Gaussianity into the error bars
for that one parameter; but if $\ln A_0$ is not part of the parameter set, marginalization distributes these effects among
the amplitude-like parameters. \citet{TakadaJain2009} showed that marginalization can reduce the impact of
non-Gaussianity; what our analysis demonstrates is the mechanism by which it does so. In particular, it is the form of the power spectrum
covariance matrix which pushes the effects from the other parameters into the amplitude-like parameters.

\section{Results: Parameter Errors}
\label{sec-results-errors}
In calculating the ratios of marginalized error bars for our model surveys, we noted an information degeneracy of $\sigma_8$ with $w$. We reduce this degeneracy by combining the results from our three hypothetical 1-Gpc$^3$ surveys (redshifts 0, 0.5, and 1) to take advantage of the increasing importance of dark energy at low redshifts. The error bar ratios (for the same parameters and number densities as before) appear in Fig.~\ref{fig:margerr} and in Table~\ref{tab:errors}.

As predicted, non-Gaussianity has virtually no effect on the matter density parameters, whereas the effect is pronounced for the amplitude-like parameters $w$, $w_a$, and $\sigma_8$. For these parameters,
the error bars in SDSS MGS-type surveys (at high wavenumbers) can be more than double that predicted by Gaussian forecasting; even for BOSS-type samples, the error bars can exceed Gaussian predictions by over 30 per cent.

\begin{table}
  \leavevmode
  \caption{Ratios of dark energy figures of merit, Gaussian to non-Gaussian}
  \label{tab:DEFoM}
  \begin{center}
  \begin{tabular}{lccc}\hline 
							& \multicolumn{3}{c}{$k_\mathrm{max}$ [$h$Mpc$^{-1}$]} \\
Survey	& 0.1		& 0.2		& 0.5\\ \hline
Continuous						& 1.21	& 1.68	& 4.72\\
Sloan Main-type ($\overline{n} = 0.003h^3$ Mpc$^{-3}$)			& 1.20	& 1.59	& 3.61\\ 
BOSS-type ($\overline{n} = 0.0003h^3$ Mpc$^{-3}$)				& 1.14	& 1.28	& 1.72\\
Euclid BAO component									& 1.10	& 1.27	& 1.89\\
\hline\hline
\end{tabular}
\end{center}
Ratio of Dark Energy Task Force figures of merit from Gaussian statistics to those from non-Gaussian. The top three rows refer to
the combination of data from three hypothetical 1-Gpc$^3$ surveys (at $z=0$, 0.5, 1). The first row shows the limiting case of a
continuous matter field; the second and third add shot noise roughly equivalent to that in the Sloan MGS
($\overline{n}=0.003h^3$ Mpc$^{-3}$) and that in BOSS ($\overline{n}=0.0003h^3$ Mpc$^{-3}$).  For this table only,
we include a galaxy bias of 1.5 for the MGS-type sample and 1.7 for the BOSS-type sample. (Thus the second and third
rows do not match the corresponding curves in Fig.~\ref{fig:DEFoM}, which include no bias). See text for
details on our modeling of the Euclid BAO component, shown in the fourth row.
\end{table}

\begin{figure}
    \leavevmode\epsfxsize=8.5cm\epsfbox{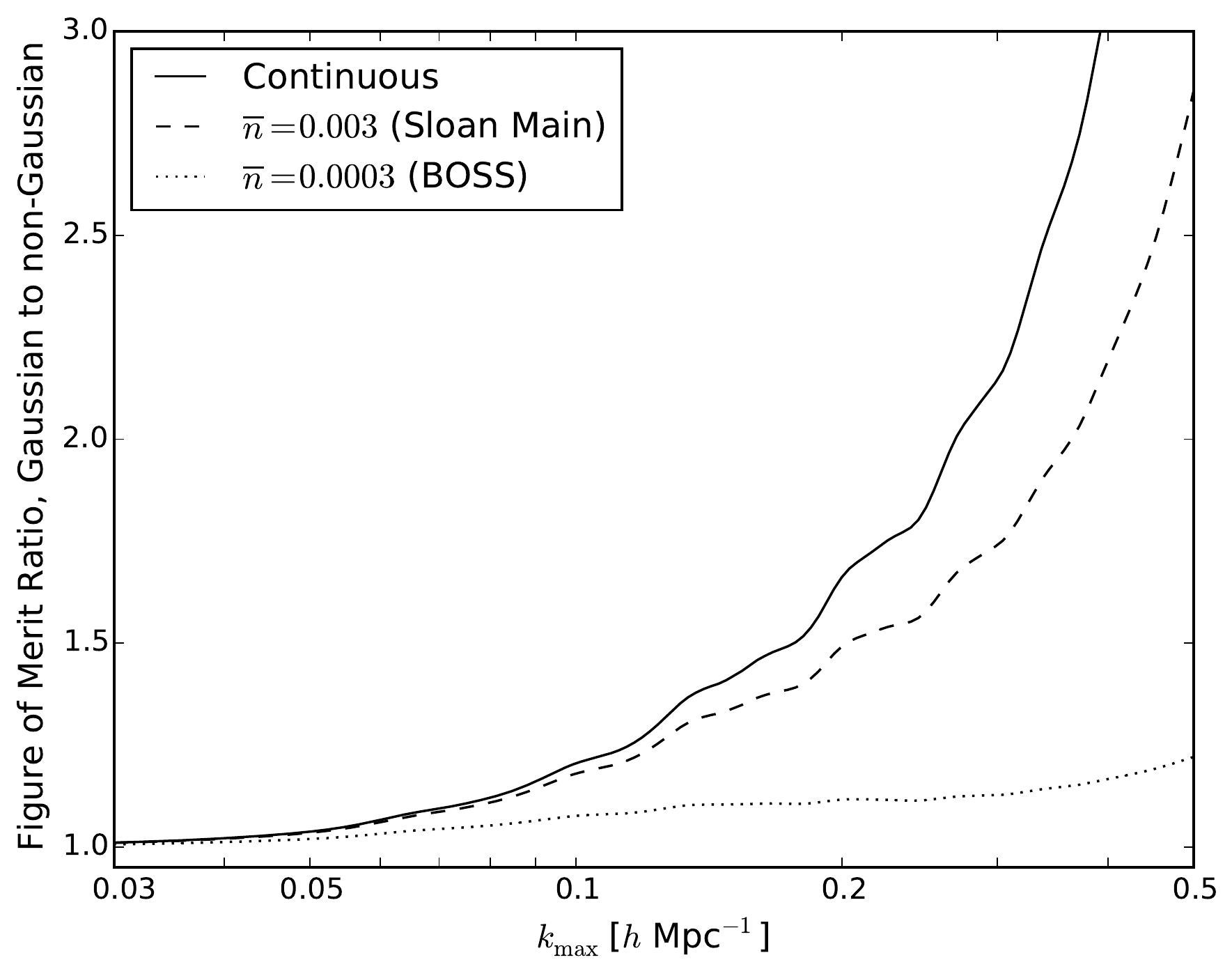}
    \caption{Ratio of the Dark Energy Task Force figure of merit under Gaussian assumptions to that under non-Gaussian assumptions. Calculations assume the combination of three hypothetical 1-Gpc$^3$ surveys at $z=0, 0.5$, and 1. The solid curve represents the limiting case of a continuous matter field; the dashed and dotted lines assume Poisson sampling with $\overline{n} = 0.003$ and 0.0003$h^3$ Mpc$^{-3}$ (comparable to the SDSS MGS and to the BOSS sample). Note that this figure differs from Table~\ref{tab:DEFoM} by including no bias for the galaxy-sampling curves.}
\label{fig:DEFoM}
\end{figure}

Focusing specifically on $w$ and $w_a$, Fig.~\ref{fig:DEFoM} and Table~\ref{tab:DEFoM} demonstrate the impact of non-Gaussianity on the Dark Energy Task Force figure of merit, defined by \citet{DEFOM} as
\begin{equation}
  \mathrm{FOM}_\mathrm{DETF} = \left[{\mathrm{det\:}}\mathrm{Cov}(w, w_a)\right]^{-1/2},
\end{equation}
where $\mathrm{Cov}(w, w_a)$ indicates the covariance submatrix for $w$ and $w_a$. To illustrate the impact of galaxy bias
(which we have ignored for the majority of this paper) we include the following biases in Table~\ref{tab:DEFoM} only but \emph{not} in
Fig.~\ref{fig:DEFoM}: to more closely simulate the surveys, we assume (in Table~\ref{tab:DEFoM}) a bias of 1.5 for the Sloan-MGS type
survey \citep{Howlett2015} 
and a bias of 1.7 for the BOSS-type survey \citep{Gil-Marin2015}. Note once again the significant effect of assuming Gaussianity: for
$k_\mathrm{max}=0.5h$ Mpc$^{-1}$ and Sloan-MGS type surveys (assuming a bias of unity), one obtains a figure of merit 2.86 times
greater than it should be. Adding the bias makes the results even more dramatic: these calculations show that even a sparse
BOSS-type survey would have a figure of merit 72 per cent higher under Gaussian assumptions than under non-Gaussian, and Gaussian assumptions would inflate the figure of merit for an MGS-type survey by 3.6 times.

In connection with this result, note first that our covariance matrix (from which we calculate the figure of merit) employs all the information in the power spectrum rather than considering the peak position only. Second, while galaxy biasing (like shot noise) can reduce the information in the spectrum, comparison of Table~\ref{tab:DEFoM} with Fig.~\ref{fig:DEFoM} demonstrates that the magnitude of this effect depends on one's assumptions about the Gaussianity of the field. Since the bias is degenerate with $\sigma_8$, it is likely that this effect is more pronounced for amplitude-like parameters. A joint analysis with the CMB could help to alleviate this degeneracy.

We emphasize in particular that it is the matter power spectrum which forms the basis for these
calculations, and that we simulate galaxies by means of shot noise. Thus it is likely that the behavior of
real galaxies will differ to some degree from that presented here.

In this connection, we must also briefly consider the following effect which we have neglected throughout this work: for very sparse samples on very small scales, our shot noise approximation 
(using $1/\overline{n}$) is no longer accurate; instead, the ratio of contributed information per $k$-mode under Gaussian and 
non-Gaussian assumptions approaches unity. This effect occurs when there is one galaxy (or fewer) per mode, in which case additional
modes would contribute no additional information. As a result one obtains a freezing of the ratios at such scales. For BOSS-type
surveys one reaches this scale around $k=0.3h$ Mpc$^{-1}$; thus for such surveys and for
amplitude-like parameters, the information ratio would freeze around 1.1 (or 1.3 including the bias). For Sloan-MGS type surveys, however, the corresponding threshold lies well above $k_\mathrm{max} = 0.5h$
Mpc$^{-1}$ indicating that the forecasting errors will indeed be as grave as shown in Fig.~\ref{fig:margerr}, Fig.~\ref{fig:DEFoM},
and Table~\ref{tab:DEFoM}.

The hypothetical surveys considered so far are artificial in that they assume the same number density and survey volume at $z=0$ as
at $z=1$. Thus we conclude by analyzing the BAO component of the upcoming Euclid survey\footnote{http://sci.esa.int/euclid/}, calculating
and combining the Fisher information in seven redshift bins ($\Delta z = 0.2$) from $z=0.65$
to $z=2.05$. Fig.~3.2 in \citet{EuclidRedBook} provides the number of galaxies in each bin, and the survey area of 15,000 deg$^2$ then yields the volume and average galaxy number density for each. The number densities in the bins range from $\overline{n}=0.0015h^3$ Mpc$^{-3}$ in the lowest bin
to 0.0001 in the highest, peaking at 0.0017 at $z \sim 1$. We include a linear
bias $b=1.5$ (similar to that of the Sloan MGS). And we find (Table~\ref{tab:DEFoM}) that Gaussian assumptions produce a figure of merit
(for $k_\mathrm{max}=0.5h$ Mpc$^{-1}$) that is almost twice as large as under non-Gaussian assumptions.

We thus conclude that forecasts in the linear regime can assume Gaussian statistics with relative impunity; however, for higher
wavenumbers it is essential to account for non-Gaussianity in surveys that seek to constrain amplitude-like parameters in general and dark
energy parameters in particular.

\section{Conclusion}
\label{sec-concl}
Forecasts of galaxy surveys' effectiveness typically assume a Gaussian field, whereas the actual field is decidedly non-Gaussian. Starting with the CWS15 approximation (Equation~\ref{eqn:CWS15}) for
the power spectrum covariance, we have developed a forecasting method
which takes into account the non-Gaussianity of the galaxy field. Our method produces results
comparable to those obtained from simulations, which is not surprising given the derivation of the CWS15 approximation itself from simulations.

Upon application of our method to hypothetical surveys, we find that the effects of non-Gaussianity are fairly minimal in the linear regime; thus Gaussian-based forecasting is relatively accurate in the regime to which it has heretofore been applied. However, we show that
non-Gaussianity can have a profound effect in the non-linear regime. For a galaxy number density of $\overline{n}=0.003h^3$ Mpc$^{-3}$
(typical of the Sloan Main Galaxy Sample) and at $k_\mathrm{max}=0.5h$ Mpc$^{-1}$, Gaussian assumptions can result in significant
overestimates of information on all parameters, up to an order of magnitude for $w$ and $\sigma_8$.

However, we find that marginalization also plays a key role in determining which parameters are
affected by non-Gaussianity. In particular, marginalization concentrates the effects into amplitude-like
parameters such as $w$, $w_a$, and $\sigma_8$. We have shown that this behavior is a consequence of the specific form of the covariance matrix. And we proceed to show that error predictions for
amplitude-like parameters can be 2.4 times as large as Gaussian forecasting would suggest. Indeed, when we take galaxy bias into account, Gaussian forecasting can produce
a dark energy figure of merit well over three times greater than warranted. And for the BAO component
of the Euclid survey in particular, we have shown that the assumption of Gaussianity would lead to figures of merit
almost twice as optimistic as they should be.

So far we have applied this method to hypothetical galaxy surveys and to Euclid. However, nothing would hinder its application to other surveys (given those surveys' parameters),
and for this purpose we make available\footnote{https://github.com/ARepp/Fisher} our code implementing this method.

We conclude that in the translinear regime for $w$, $w_a$, and $\sigma_8$,  accuracy in forecasting depends critically
on one's assumptions about the statistics of the underlying field, and that the assumption of Gaussianity can lead to profoundly
erroneous forecasts.

This sensitivity to non-Gaussianity is a consequence of a fundamental limitation of the usual $\delta$-based power spectrum,
namely, that the information content saturates at high $k$-values, rendering the remaining information inaccessible.
\citet{RimesHamilton2005} and \citet{Neyrinck2006} note that consideration of higher-order statistics
does not alleviate this problem, a fact
which \citet{Wolk2013} confirmed empirically with Canada-France-Hawaii Telescope Legacy Survey (CFHTLS) data.

\citet{Neyrinck2009} demonstrate that a logarithmic transformation mitigates this issue. \citet{Astar0, Astar} rederive and then refine this result, obtaining the alternate ``sufficient statistics'' observable $A^*$, which circumvents the information plateau by transforming
away the non-Gaussian effects. \citet{WolkForecast} show that the use of $A^*$
can increase the constraints on cosmological parameters by up to a factor of two, and \citet{WolkAnalyA} provide
an analytic framework for $A^*$-based forecasting. In addition, \citet{WolkNeutrino} show that the combination of $A^*$ with
the CWS15 covariance matrix (Equation~\ref{eqn:CWS15}) can improve constraints on neutrino masses by almost a factor of three,
compared to using the matter power spectrum. Thus, non-Gaussianity will have a profound impact on forecasts of future surveys,
if those forecasts use the traditional $\delta$ power spectrum. However, an alternate statistic for obtaining accurate forecasts is already available.
\\
\\
IS, JC, and MW acknowledge NASA grants NNX12AF83G and NNX10AD53G for support. The research
leading to these results has received funding from the European
Research Council under the European Union's Seventh Framework Programme
(FP/2007-2013) / ERC Grant Agreement No. [616170]. We also thank Mark Neyrinck for
his helpful suggestions for improvement of the manuscript.

\bibliographystyle{mn2e}
\bibliography{NonGaussianForecasting}

\appendix
\section{Marginalization and the Non-Linear Amplitude Parameter}
\label{sec:appendix}
We here demonstrate that if the non-linear amplitude parameter $\ln A_0$ is one of the parameters under consideration (and if the Fisher matrix is invertible), then non-Gaussianity affects only the covariance of $\ln A_0$. In this derivation, a superscript $G$ denotes a quantity calculated under the assumption of Gaussianity.

We begin with Equation~\ref{eqn:SSF}:
\begin{equation}
F_{\alpha\beta} = F^G_{\alpha\beta} - \sigma^2_\mathrm{min} \frac{F^G_{\alpha \, \ln\!A_0} F^G_{\ln\!A_0\,\beta}}{1+\sigma^2_\mathrm{min} F^G_{\ln\!A_0\,\ln\!A_0}}.
\nonumber
\end{equation}
The Sherman-Morrison formula states that for an invertible square matrix $A$ and vectors $u$ and $v$,
\begin{equation}
(A+uv^T)^{-1} = A^{-1} - \frac{A^{-1}uv^T A^{-1}}{1 + v^T A^{-1} u}.
\end{equation}
We apply this formula by setting $A=F^G$ and
\begin{equation}
u_\alpha = -v_\alpha = \frac{\sigma_\mathrm{min} F^G_{\ln\!A_0 \, \alpha}}{\sqrt{1+\sigma^2_\mathrm{min} F^G_{\ln\!A_0 \, \ln\!A_0}}}.
\end{equation}
For notational convenience we assume summation over indices appearing as both super- and
subscripts (without thereby implying any contra/covariance). The result is as follows:
\begin{eqnarray}
\lefteqn{F^{-1}_{\alpha\beta} = (F^G)^{-1}_{\alpha\beta}}\\
   &  & + \frac{\sigma^2_\mathrm{min} \left(\rule{0pt}{10pt}(F^G)^{-1}_{\alpha\gamma} (F^G)^\gamma_{\ln \!A_0}\right) \left((F^G)^{-1}_{\beta\gamma} (F^G)^\gamma_{\ln\!A_0}\right)}{1+\sigma^2_\mathrm{min}\left( F^G_{\ln\!A_0\,\ln\!A_0} - (F^G)^\gamma_{\ln\!A_0} (F^G)^{-1}_{\gamma\delta} (F^G)^\delta_{\ln\!A_0}\right)}\nonumber.
\end{eqnarray}

Now if the amplitude parameter $\ln A_0$ is part of the
parameter set, then by definition
\begin{equation}
(F^G)^{-1}_{\alpha\delta} (F^G)^\delta_{\ln\!A_0} = \delta_{\alpha\,\ln\!A_0}.
\end{equation}
It follows that
\begin{equation}
F^{-1}_{\alpha\beta} = (F^G)^{-1}_{\alpha\beta} + \sigma^2_\mathrm{min} \: \delta_{\alpha \, \ln\!A_0}\:\delta_{\ln \!A_0\, \beta},
\end{equation}
so that the marginalized errors are
\begin{equation}
\sigma^2_\alpha = \left\{ \begin{array}{ll}
				(\sigma^2)^G_\alpha	 + \sigma^2_\mathrm{min}	&	\mbox{if $\alpha = \ln A_0$} \\
				(\sigma^2)^G_\alpha							&      \mbox{if $\alpha \neq \ln A_0$}
                             \end{array} \right.,
\end{equation}
which was to be proved.

\label{lastpage}
\end{document}